\documentclass[runningheads]{llncs}
\usepackage{graphicx}
\usepackage{cite}
\usepackage{graphicx}
\usepackage{amsmath}
\usepackage{yhmath} 

\usepackage{amsfonts} 
\usepackage{amssymb}
\usepackage{color}
\usepackage{latexsym}

\usepackage{algorithm}
\usepackage[noend]{algorithmic}
\usepackage{multirow} 
\usepackage{amsmath}

\newtheorem{observation}[theorem]{Observation}
\newcommand{\old}[1]{{}}

\newcommand{\B}{{{\cal{B}}}}
\renewcommand{\P}{{{\mathcal{P}}}}
\newcommand{\T}{{{\cal{T}}}}

\newcommand{\ceil}[1]{\left\lceil {{#1}} \right\rceil}

\newcommand{\A}{{{\textsc{All}}}}
\newcommand{\AL}{{{\textsc{All-L}}}}
\newcommand{\AR}{{{\textsc{All-R}}}}
\newcommand{\ALR}{{{\textsc{All-LR}}}}
\newcommand{\LM}{{{\textsc{LeftMost}}}}
\newcommand{\RM}{{{\textsc{RightMost}}}}
\newcommand{\ML}{{{\textsc{Ml}}}}
\newcommand{\MR}{{{\textsc{Mr}}}}
\newcommand{\MLR}{{{\textsc{Mlr}}}}

\newcommand{\Miss}{{{\textsc{All-1}}}}
\newcommand{\MissL}{{{\textsc{All-1-L}}}}
\newcommand{\MissR}{{{\textsc{All-1-R}}}}
\newcommand{\MissLR}{{{\textsc{All-1-LR}}}}

\begin{document}

\title{Dynamic Euclidean Bottleneck Matching}
%\thanks{This work was partially supported by Grant 2016116 from the United States -- Israel Binational Science Foundation.}}

\author{A.\,Karim Abu-Affash\inst{1} \and
Sujoy Bhore\inst{2} \and
Paz Carmi\inst{3} 
}
\authorrunning{A. K. Abu-Affash, S. Bhore, P. Carmi}
\institute{Department of Software Engineering, Shamoon College of Engineering, Israel \\ \email{abuaa1@sce.ac.il} \and
Department of Computer Science, Indian Institute of Technology Bombay, India \email{sujoy@cse.iitb.ac.in} \and
Computer Science Department, Ben-Gurion University, Israel \\ \email{carmip@cs.bgu.ac.il} 
}

\maketitle

\begin{abstract}
%We present an algorithm for computing a fully dynamic Euclidean bottleneck matching for a set of points in the plane. 
A fundamental question in computational geometry is for a set of input points in the Euclidean space, that is subject to discrete changes (insertion/deletion of points at each time step), whether it is possible to maintain an approximate bottleneck matching in sublinear update time. In this work, we answer this question in the affirmative for points on a real line and for points in the plane with a bounded geometric spread. 

For a set $P$ of $n$ points on a line, we show that there exists a dynamic algorithm that maintains a bottleneck matching of $P$ and supports insertion and
deletion in $O(\log n)$ time. Moreover, we show that a modified version of this algorithm maintains a minimum-weight matching with $O(\log n)$ update (insertion and deletion) time. Next, for a set $P$ of $n$ points in the plane, we show that a ($6\sqrt{2}$)-factor approximate bottleneck matching of $P_k$, at each time step $k$, can be maintained in $O(\log{\Delta})$ amortized time per insertion and $O(\log{\Delta} + |P_k|)$ amortized time per deletion, where $\Delta$ is the geometric spread of $P$.

\end{abstract}

\section{Introduction}
Let $P$ be a set of $n$ points in the plane. Let $G=(P,E)$ denote the complete graph over $P$, which is an undirected weighted graph with $P$ as the set of vertices and the weight of every edge $(p,q)\in E$ is the Euclidean distance $|pq|$ between $p$ and $q$. For a perfect matching $M$ in $G$, let $bn(M)$ be the length of the longest edge. A perfect matching $M^*$ is called a bottleneck matching of $P$, if for any other perfect matching $M$, $bn(M)\ge bn(M^*)$.

Computing Euclidean bottleneck matching was studied by Chang et al.~\cite{Chang92}. They proved that such kind of matching is a subset of $17$-RNG (relative neighborhood graph) and presented  an $O(n^{3/2}\log^{1/2} n)$-time algorithm to compute a bottleneck matching. 
%In subsequent work, Su and Chang~\cite{su1991computing} gave an algorithm with improved running time. 
In fact, a major caveat of the Euclidean bottleneck matching algorithms was that they relied on Gabow and Tarjan~\cite{gabow1988algorithms} as an initial step (as also noted by Katz and Sharir~\cite{katz2022bottleneck}). 
In recent work, Katz and Sharir~\cite{katz2022bottleneck} showed that the Euclidean bottleneck matching for a set of $n$ points in the plane can be computed in $O(n^{\omega/2}\log n)$ deterministic time, where $\omega\approx 2.37$ is the exponent of matrix multiplication.
For general graphs of $n$ vertices and $m$ edges, Gabow and Tarjan~\cite{gabow1988algorithms} gave an algorithm for maximum bottleneck matching that runs in $O(n^{5/2}\sqrt{\log n})$ time. 
Bottleneck matchings were also studied for points in higher dimensions and in other metric spaces~\cite{EfratK00}, with non-crossing constraints~\cite{Abu-Affash14, Abu-Affash15}, and on multichromatic instances~\cite{Abu-AffashBC20}.

In many applications, the input instance changes over a period of time, and the typical objective is to build dynamic data structures that can update solutions efficiently rather than computing everything from  scratch. In recent years, several dynamic algorithms were designed for geometric optimization problems; see~\cite{agarwal2022dynamic, bhore2020dynamic, BBCCI21, bhore2022algorithmic, chan2010dynamic}. 
Motivated by this, we study the bottleneck matching for dynamic point set in the Euclidean plane. 
In our setting, the input is a set of points in the Euclidean plane 
and the goal is to devise a dynamic algorithm that maintains a bottleneck matching of the points and supports dynamic changing of the input due to insertions and deletions of points. Upon a modification to the input, the dynamic algorithm should efficiently update the bottleneck matching of the new set.

\subsection{Related Work}\label{related work}
Euclidean matchings have been a major subject of investigation for several decades due to their wide range of applications in operations research, pattern recognition, statistics, robotics, and VLSI; see~\cite{cho2018indoor, MarcotteS91}. The Euclidean minimum-weight matching, where the objective is to compute a perfect matching with the minimum total weight, was studied by Vaidya~\cite{vaidya1989geometry} who gave the first sub-cubic algorithm ($O(n^{5/2}\log^4 n)$) by exploiting geometric structures. Varadrajan~\cite{varadarajan1998divide} presented an $O(n^{3/2}\log^5 n)$-time algorithm for computing a minimum-weight matching in the plane, which is the best-known running time for Euclidean minimum-weight matching till date. Agarwal et al.~\cite{agarwal2000vertical} gave a near quadratic time algorithm for the bipartite version of the problem, improving upon the sub-cubic algorithm of Vaidya~\cite{vaidya1989geometry}. Several recent approximation algorithms were developed with improved running times for bipartite and non-bipartite versions; see~\cite{AgarwalCRX22, RaghvendraA20, AgarwalV04}. 

%%%%%%%%%
\old{
\paragraph{Dynamic Matching.} 
%In many applications, the input instance changes over a period of time, and the typical objective is to build dynamic data structures that can update solutions efficiently rather than computing everything from the scratch. In recent years, several dynamic algorithms were designed for geometric optimization problems; see~\cite{agarwal2022dynamic, bhore2020dynamic, bhore2022algorithmic, chan2010dynamic}. 
%Motivated by this, we study the bottleneck matching for dynamic point set in the Euclidean plane. 
%In our setting, the input is a set of points in the Euclidean plane that is subject to dynamic updates, such as an insertion or a deletion of a point at every time step, and the objective is to maintain an approximate bottleneck matching at each step. 
Dynamic matchings have been studied extensively on graphs in recent years (see Section~\ref{related work} for a detailed overview), and the best-known result requires linear update time on the number of edges to maintain a $(1+\epsilon)$-approximate solution. 
In the Euclidean setting, where the underlying graph is a complete graph, the objective is an approximate matching with sublinear update time. 
However, note that there exists a simple dynamic sequence of input points in the plane such that it is not possible to maintain an approximate bottleneck matching in sublinear update time. Consider the following adversarial sequence:
In the first $j-1$ steps, points are inserted on a line in a sequential manner. At each step, a pair of points is introduced such that the points are $\epsilon > 0$ distance apart, for an infinitesimally small constant $\epsilon$. Moreover, the newly introduced pair is exactly distance $1$ apart from the previous pair. 
Now, at step~$j$, two points $p_j$ and $p'_j$ are inserted such that each of them is vertically $1$ length below of the leftmost and rightmost pair; see Figure~\ref{fig:seq}. While the optimum bottleneck at step~$j$ is $1$, however, this could  incur a linear number of changes for any algorithm to maintain such matching. 

\begin{figure}
    \centering
    \includegraphics{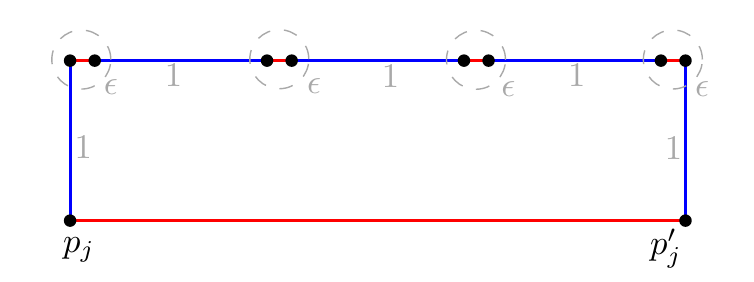}
    \caption{Blue matching is the optimum bottleneck matching. Red matching is the enforced matching with an arbitrarily large bottleneck unless an algorithm makes asymptotically linear changes.}
    \label{fig:seq}
\end{figure}

%\begin{itemize}
%\item Model. 
%\item State-of-the-art. 
%\item Lower bound example. 
%\item Pose the ques. on sublinear algorithms.
%\item Connect with the dynamic graph matching. 
%\end{itemize}

%In the incremental setting, where the edges are only allowed to be inserted but not deleted, Gupta~\cite{gupta2014maintaining} gave a $(1+\varepsilon)$-approximation for
%this setting, with amortized $O(\log^2 n)$-update time. Later, Grandoni et al.~\cite{solomon2016fully} presented a deterministic algorithm that maintains a $(1+\varepsilon)$-approximate matching with constant amortized update time per insertion.

} % end old

\paragraph{Dynamic Graph Matching.} In this problem, the objective is to maintain a maximal cardinality matching as the input graph is subject to discrete changes, i.e., at each time step, either a vertex (or edge) is added or deleted.
Dynamic graph matching algorithms have been extensively studied over the past few decades. However, most of these algorithms consider dynamic graphs which are subject to discrete edge updates, as also noted by Grandoni et al.~\cite{solomon2016fully}. 
Sankowski~\cite{sankowski2007faster} showed how to maintain the size of the maximum matching with $O(n^{1.495})$ worst-case update time. Moreover, it is known that maintaining an exact matching requires polynomial update time under complexity conjectures~\cite{abboud2014popular}. Therefore, most of the research has been focused on maintaining an approximate solution. 
It is possible to maintain a $2$-approximate matching with constant amortized update time~\cite{solomon2016fully}. However, one can maintain a $(1+\varepsilon)$-approximate solution in the fully-dynamic setting with
update time $O(\sqrt{m}/\varepsilon^2)$~\cite{gupta2013fully}. 

\paragraph{Online Matching.} 
Karp, Vazirani, and Vazirani studied the bipartite vertex-arrival model in their seminal work~\cite{karp1990optimal}. 
Most of the classical online matching algorithms are on the server-client paradigm, where one side of a bipartite graph is revealed at the beginning. Raghvendra~\cite{raghvendra2018optimal} studied the online bipartite matching problem for a set of points on a line (see also~\cite{koutsoupias2003online}). Gamlath et al.~\cite{gamlath2019online} studied the online matching problem on edge arrival model. Despite of the remarkable progress of the online matching problem over the decades, the online minimum matching with vertex arrivals has not been studied (where no side is revealed at the beginning). 
%Note that, in this case, the cost of an optimum solution in subsequent steps neither monotonically increases or decreases, which makes it theoretically rather challenging to establish the bounds on the competitive ratio. 

\subsection{Our contribution}
\old{
In this paper, we prove the following two theorems:
\begin{theorem} \label{thm:Theorem1}
Let $P$ be a set of $n$ points on a line. There exists a dynamic algorithm that maintains a bottleneck matching of $P$ and supports insertion and deletion in $O(\log n)$ time.
\end{theorem}
\begin{theorem} \label{thm:Theorem2}
Let $P$ be a set of points in the plane.
There exists a dynamic algorithm that maintains a $(6\sqrt{2})$-approximate bottleneck matching of $P_k$, at each time step $k$, and supports insertion in $O(\log{\Delta})$ amortized time and deletion in $O(\log{\Delta} + |P_k|)$ amortized time, where $\Delta$ is the geometric spread of $P$. 
\end{theorem}
}
In Section~\ref{Sec:Dynmic1D}, we present a dynamic algorithm that maintains a bottleneck matching of a set $P$ of $n$ points on a line with $O(\log n)$ update (insertion or deletion) time. Then, in Section~\ref{sec:Ext}, we generalize this algorithm to maintain a minimum-weight matching of $P$ with $O(\log n)$ update time. For a set $P$ of points in the plane with bounded geometric spread $\Delta$, in Section~\ref{Sec:Dynmic2D}, we present a dynamic algorithm that maintains a $(6\sqrt{2})$-approximate bottleneck matching of $P_k$, at each time step $k$, and supports insertion in $O(\log{\Delta})$ amortized time and deletion in $O(\log{\Delta} + |P_k|)$ amortized time.

%%%%%%%%%%%%%%%%%%%%%%%%%%%%%%%%%%%%%%%%%%%%%%%%%%%%%
%%%%%%%%%%%%%%%%%%%%%%%%%%%%%%%%%%%%%%%%%%%%%%%%%%%%%
\section{Dynamic Bottleneck Matching in 1D} \label{Sec:Dynmic1D}
Let $P=\{p_1,p_2,\ldots,p_n\}$ be a set of $n$ points located on a horizontal line, such that $p_i$ is to the left of $p_{i+1}$, for every $0 \le i < n$.
In this section, we present a dynamic algorithm that maintains a bottleneck matching of $P$ with logarithmic update time. 
Throughout this section, we assume that $n$ is even and two points are added or deleted in each step. However, our algorithm can be generalized for every $n$ and every constant number of points added or deleted in each step, regardless of the parity of $n$; see Section~\ref{sec:Ext}.

\begin{figure}[htb]
    \centering
    \includegraphics[width=0.8\textwidth]{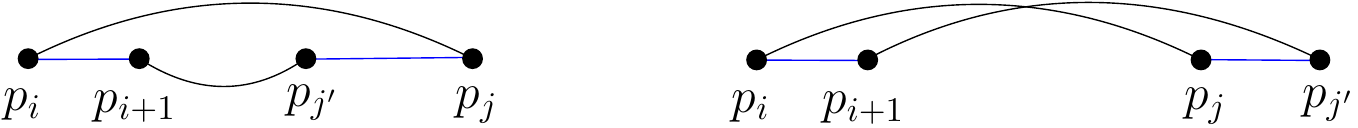}
    \caption{The matching of the points $\{p_i,p_{i+1},p_j,p_{j'}\}$ in $M'$ (in black) and in $M''$ (in blue).}
    \label{fig:obs}
\end{figure}

\begin{observation}\label{obs:sec-1}
There exists a bottleneck matching $M$ of $P$, such that each point $p_i\in P$ is matched to a point from $\{p_{i-1},p_{i+1}\}$.
\end{observation}
\begin{proof}
Let $M'$ be a bottleneck matching of $P$ in which there exists at least one point $p_i$ that is not matched to $p_{i-1}$ or to $p_{i+1}$. We do the following for each such a point $p_i$.
Let $p_i$ be the leftmost point in $P$ that is matched in $M'$ to a point $p_j$, where $j>i+1$.
Let $p_{j'}$ be the point that is matched to $p_{i+1}$, and notice that $j' > i+1$.
Let $M''$ be the matching obtained by replacing the edges $(p_i,p_j)$ and $(p_{i+1},p_{j'})$ in $M'$ by the edges $(p_i,p_{i+1})$ and $(p_{j},p_{j'})$; see Figure~\ref{fig:obs}.
Clearly, $|p_ip_{i+1}| \le |p_ip_j|$ and $|p_jp_{j'}| \le \max\{ |p_ip_j|,|p_{i+1}p_{j'}|\}$.
Therefore, $M''$ is also a bottleneck matching in which $p_i$ is matched to $p_{i+1}$. 
\end{proof}

Throughout the rest of this section, we refer to the bottleneck matching that satisfies Observation~\ref{obs:sec-1} as the optimal matching, and notice that this matching is unique.

%%%%%%%%%%%%%%%%%%%%%%%%%%%%%%%%%%%%%%%%%%%%%%%%%%%%%%%%%%%%%%%%%%%%
\subsection{Preprocessing} \label{sec:1DPreprocessing}
Let $M$ be the optimal matching of $P$ and let $bn(M)$ denote its bottleneck. 
Clearly, $M$ can be computed in $O(n)$ time.
We maintain $M$ in a full AVL tree $\T$, such that the leaves of $\T$ are the points of $P$, and each intermediate node has exactly two children and contains some extra information, propagated from its children.
For a node $v$ in $\T$, let $T_v$ be the sub-tree of $\T$ rooted at $v$, and let $P_v$ be the subset of $P$ containing the points in the leaves of $T_v$.
For each node $v$ in $\T$, let $lc(v), rc(v)$ be the left and the right children of $v$, respectively, and $p(v)$ be the parent of $v$.
%; see Figure~\ref{fig:dy-mat-1}.

Each node $v$ in $\T$ contains the following seven attributes about the optimal matching of the points in $P_v$: 
\begin{enumerate}
    %\vspace{-.05cm}
    \item \LM$(v)$ - the leftmost point in $P_v$.
    %\vspace{.05cm}
    \item \RM$(v)$ - the rightmost point in $P_v$.
    %\vspace{.05cm}
    \item $\pi(v) = |\RM(lc(v))\LM(rc(v)|$ - the Euclidean distance \\ between $\RM(lc(v))$ and $\RM(lc(v))$.
    %\vspace{.05cm}
    \item \A$(v)$ - cost of the matching of the points in $P_v$.
    %\vspace{.05cm}
    \item \AL$(v)$ - cost of the matching of the points in $P_v\setminus\{\LM(v)\}$.
    %\vspace{.05cm}
    \item \AR$(v)$ - cost of the matching of the points in $P_v\setminus\{\RM(v)\}$.
    %\vspace{.05cm}
    \item \ALR$(v)$ - cost of the matching of the points in $P_v\setminus\{\LM(v),$\\$\RM(v)\}$.
\end{enumerate}

Now, we describe how to compute the values of the attributes in each node $v$. The computation is bottom-up. That is, we first initialize the attributes of the leaves and then, for each intermediate node $v$, we compute its attributes from the attributes of its children $lc(v)$ and $rc(v)$.  %Let $\pi(v) = |\RM(lc(v))\LM(rc(v)|$, $|pq|$ is the Euclidean distance between $p$ and $q$ %Let $p=\LM(rc(v))$ and $p=\RM(lc(v))$.

For each leaf $v$ in $\T$, we set $\A(v)$ and $\ALR(v)$ to be $\infty$, $\AL(v)$ and $\AR(v)$ to be 0, and $\LM(v)$ and $\RM(v)$ to be $v$.
For each intermediate $v$ in $\T$, we compute its attributes as follows.

\begin{align*}
\A(v) \leftarrow &  \min\Big\{ \max\big\{\A(lc(v)) \,, \A(rc(v)) \big\}  \,, \\
& \quad \ \quad \ \max\big\{\AR(lc(v)) \,, \AL(rc(v))  \,, \pi(v) \big\}\Big\}\,.\\ 
\end{align*}
\begin{align*}
\AL(v) \leftarrow &  \min\Big\{ \max\big\{\AL(lc(v)) \,, \A(rc(v)) \big\}  \,, \\
& \quad \ \quad \ \max\big\{\ALR(lc(v)) \,, \AL(rc(v))  \,, \pi(v) \big\}\Big\}\,.\\ 
\AR(v) \leftarrow &  \min\Big\{ \max\big\{\A(lc(v)) \,, \AR(rc(v)) \big\}  \,, \\
& \quad \ \quad \ \max\big\{\AR(lc(v)) \,, \ALR(rc(v))  \,, \pi(v) \big\}\Big\}\,.\\
\ALR(v) \leftarrow &  \min\Big\{ \max\big\{\AL(lc(v)) \,, \AR(rc(v)) \big\}  \,, \\
& \quad \ \quad \ \max\big\{\ALR(lc(v)) \,, \ALR(rc(v))  \,, \pi(v) \big\}\Big\}\,.\\ 
\end{align*}
Clearly, these values can be computed in constant time, for each node $v$ in $\T$, given the attributes of its children.
Therefore, the preprocessing time is $O(n)$. 
%Assume now that $\mathcal{T}_i$ is the tree with all the above information at step~$i$, for some $i\in \mathbb{N}$. Now, we describe the dynamic update operation that can be performed in poly-logarithmic update time.

\begin{lemma} \label{lemma:1Dcorrect}
Let $r^*$ be the root of $\T$. Then, $\A(r^*)=bn(M)$.
\end{lemma}
\begin{proof}
For a node $v$ in $\T$ where $|P_v|$ is even, let $M_v$ denote the optimal matching
%~\footnote{Recall that this matching is satisfying Observation~\ref{obs:sec-1}}
of the points in $P_v$, and let $\MLR_v$ denote the optimal matching of the points in $P_v\setminus\{\LM(v),\RM(v)\}$.
For a node $v$ in $\T$ where $|P_v|$ is odd, let $\ML_v$ denote the optimal matching of the points in $P_v\setminus\{\LM(v)\}$, and let $\MR_v$ denote the optimal matching of the points in $P_v\setminus\{\RM(v)\}$.

To prove the lemma, we prove a stronger claim. For each node $v$ in $\T$, we prove that
\begin{itemize}
    \item if $|P_v|$ is even, then $\A(v) = bn(M_v)$, $\AL(v)=\AR(v)=\infty$, and $\ALR(v)=bn(\MLR_v)$.
    \item if $|P_v|$ is odd, then $\A(v) = \ALR(v) = \infty$, $\AL(v)=bn(\ML_v)$, and $\AR(v)=bn(\MR_v)$.
\end{itemize}
The proof is by induction on the height of $v$ in $\T$. \\
\textbf{Base case:} The claim holds for each leaf $v$ in $\T$, since $|P_v|=1$ and we initialize the attributes of $v$ by the values $\A(v)=\ALR(v)=\infty$ and $\AL(v)=\AR(v)=0$. Moreover, for each node $v$ in height one, we have $|P_v|=2$ and $v$ has two leaves $l$ and $r$ at height zero. Therefore,
\begin{align*}
\A(r) = & \ \min\Big\{ \max\big\{\A(l) \,, \A(r)) \big\}  \,, \max\big\{\AR(l) \,, \AL(r)  \,, \pi(v) \big\}\Big\} \\
= & \ \min\Big\{ \max\big\{\infty \,, \infty \big\}  \,, \max\big\{0 \,, 0  \,, |lr| \big\}\Big\} =  |lr| \,. \\
\AL(v) = & \ \min\Big\{ \max\big\{\AL(l) \,, \A(r) \big\}  \,, \max\big\{\ALR(l) \,, \AL(r)  \,, \pi(v) \big\}\Big\} \\
= & \ \min\Big\{ \max\big\{\infty \,, 0 \big\}  \,, \max\big\{0 \,,\infty  \,, |lr| \big\}\Big\} = \infty \,. \\
\end{align*}
\begin{align*}
\AR(v) = & \ \min\Big\{ \max\big\{\A(l) \,, \AR(r) \big\}  \,, \\
& \ \quad \ \quad \ \max\big\{\AR(l) \,, \ALR(r)  \,, \pi(v) \big\}\Big\} \\
= & \ \min\Big\{ \max\big\{0 \,, \infty \big\}  \,, \max\big\{\infty \,, 0  \,, |lr| \big\}\Big\} = \infty \,. \\
\ALR(v) = & \ \min\Big\{ \max\big\{\AL(l) \,, \AR(r) \big\}  \,, \\
& \ \quad \ \quad \ \max\big\{\ALR(l) \,, \ALR(r)  \,, \pi(v) \big\}\Big\} \\ 
= & \ \min\Big\{ \max\big\{0 \,, 0 \big\}  \,, \max\big\{\infty \,,\infty  \,, |lr| \big\}\Big\} = 0 \,.
\end{align*}
\textbf{Induction step:} We prove the claim for each node $v$ at height $h>1$. Let $l=lc(v)$ and $r=rc(v)$.
Let $p$ and $q$ be the rightmost and the leftmost points in $P_l$ and $P_r$, respectively. Thus, $\pi(v) = |pq|$.
We distinguish between four cases. %Due to space limitation, we provide a proof for Case~1. %The proofs of the other three cases are similar and moved to Appendix~\ref{Lemma1Proof}. \\

%%%%%%%%%%%%%%%%%%%%%%%%%%%%%%%%%%%%%%%%%%%%%%%%%%%%%%%%%%%%%%%%%%%
%%  Case 1 $P_v$ is even and both $|P_l|$ and $|P_r|$ are even
%%%%%%%%%%%%%%%%%%%%%%%%%%%%%%%%%%%%%%%%%%%%%%%%%%%%%%%%%%%%%%%%%%%%
\noindent
\textbf{Case~1: \boldmath{$|P_v|$} is even and both $|P_l|$ and $|P_r|$ are even.} \\
%Then we distinguish between two cases. \\
%\textbf{Case~1.1:} $|P_l|$ and $|P_r|$ are even. 
Since $|P_v|$ is even, $M_v$ consists of the optimal matching $M_l$ of $P_l$ and the optimal matching $M_r$ of $P_r$, and $bn(M_v) = \max\{bn(M_l),bn(M_r)\}$. 
Moreover, $\MLR_v$ consists of the optimal matching $\MLR_l$ of $P_l\setminus\{\LM(l),\RM(l)\}$, the optimal matching $\MLR_r$ of $P_r\setminus\{\LM(r),\RM(r)\}$, and the edge $(p,q)$. Thus, $bn(\MLR_v) = \max\{bn(\MLR_l),bn(\MLR_r),|pq|\}$.

By the induction hypothesis, $\A(l)=bn(M_l)$, $\A(r)=bn(M_r)$, $\ALR(l) = bn(\MLR_l)$, $\ALR(r) = bn(\MLR_r)$, and $\AL(l)=\AR(l)=\AL(l)=\AR(l)=\infty$. 
Therefore, we have
\begin{align*}
 \A(r) = & \ \min\Big\{ \max\big\{\A(l) \,, \A(r)) \big\} \,, \max\big\{\AR(l) \,, \AL(r)  \,, \pi(v) \big\}\Big\} \\
= & \ \min\Big\{ \max\big\{bn(M_l) \,, bn(M_r) \big\}  \,, \max\big\{\infty \,, \infty  \,, |pq| \big\}\Big\}  \\
= & \  \max\{bn(M_l) \,, bn(M_r)\} = bn(M_v)\,. \\
\AL(v) = & \ \min\Big\{ \max\big\{\AL(l) \,, \A(r) \big\}  \,, \\
& \ \quad \ \quad \ \max\big\{\ALR(l) \,, \AL(r)  \,, \pi(v) \big\}\Big\} \\
= & \ \min\Big\{ \max\big\{\infty \,, bn(M_r) \big\}  \,, \max\big\{bn(\MLR_l)  \,, \infty \,, |pq| \big\}\Big\} = \infty \,. \\
\AR(v) = & \ \min\Big\{ \max\big\{\A(l) \,, \AR(r) \big\}  \,, \\
& \ \quad \ \quad \ \max\big\{\AR(l) \,, \ALR(r)  \,, \pi(v) \big\}\Big\} \\
= & \ \min\Big\{ \max\big\{bn(M_l) \,, \infty \big\}  \,, \max\big\{\infty \,,  bn(\MLR_r)   \,, |pq| \big\}\Big\} = \infty \,. \\
\end{align*}
\begin{align*}
\ALR(v) = & \ \min\Big\{ \max\big\{\AL(l) \,, \AR(r) \big\}  \,, \\
& \ \quad \ \quad \ \max\big\{\ALR(l) \,, \ALR(r)  \,, \pi(v) \big\}\Big\} \\ 
= & \ \min\Big\{ \max\big\{\infty  \,, \infty \big\}  \,, \max\big\{bn(\MLR_l)  \,, bn(\MLR_r)  \,, |pq| \big\}\Big\}  \\
 = & \ \max\big\{bn(\MLR_l)  \,, bn(\MLR_r)  \,, |pq| \big\} = bn(\MLR_v) \,.
\end{align*}

\noindent
\textbf{Case~2: \boldmath{$|P_v|$} is even and both $|P_l|$ and $|P_r|$ are odd.} \\
Since $|P_v|$ is even, $M_v$ consists of the optimal matching $\ML_r$ of $P_r\setminus\{\LM(r)\}$, the optimal matching $\MR_l$ of $P_l\setminus\{\RM(l)\}$, and the edge $(p,q)$. Thus, $bn(M_v) = \max\{bn(\ML_r),bn(\MR_l),|pq|\}$.
Moreover, $\MLR_v$ consists of the optimal matching $\ML_l$ of $P_l\setminus\{\LM(l)\}$ and the optimal matching $\MR_r$ of $P_r\setminus\{\RM(r)\}$, and $bn(\MLR_v) = \max\{bn(\ML_l),bn(\MR_r)\}$.

By the induction hypothesis, $\A(l)=\ALR(l)=\A(r)=\ALR(r)=\infty$, $\AR(l)=bn(\MR_l)$, $\AL(l)=bn(\ML_l)$, $\AR(r)=bn(\MR_r)$, and $\AL(r)=bn(\ML_r)$. 
%Moreover, we need to show that $\AR(v)$ and $\AL(v)$ are $\infty$. 
Therefore, we have
\begin{align*}
 \A(r) = & \ \min\Big\{ \max\big\{\A(l) \,, \A(r)) \big\}  \,, \max\big\{\AR(l) \,, \AL(r)  \,, \pi(v) \big\}\Big\} \\
= & \ \min\Big\{ \max\big\{\infty \,, \infty \big\}  \,, \max\big\{bn(\MR_l) \,, bn(\ML_r)  \,, |pq| \big\}\Big\}  \\
= & \ \max\big\{bn(\MR_l) \,,bn(\ML_r) \,, |pq| \big\} = bn(M_v) \,. \\
\AL(v) = & \ \min\Big\{ \max\big\{\AL(l) \,, \A(r) \big\}  \,, \\
& \ \quad \ \quad \ \max\big\{\ALR(l) \,, \AL(r)  \,, \pi(v) \big\}\Big\} \\
= & \ \min\Big\{ \max\big\{bn(\ML_l) \,, \infty \big\}  \,, \max\big\{\infty  \,, bn(\ML_r) \,, |pq| \big\}\Big\} = \infty \,. \\
\AR(v) = & \ \min\Big\{ \max\big\{\A(l) \,, \AR(r) \big\}  \,, \\
& \ \quad \ \quad \ \max\big\{\AR(l) \,, \ALR(r)  \,, \pi(v) \big\}\Big\} \\
= & \ \min\Big\{ \max\big\{\infty \,, bn(\MR_r) \big\}  \,, \max\big\{bn(\MR_l) \,, \infty  \,, |pq| \big\}\Big\} = \infty \,. \\
\ALR(v) = & \ \min\Big\{ \max\big\{\AL(l) \,, \AR(r) \big\}  \,, \\
& \ \quad \ \quad \ \max\big\{\ALR(l) \,, \ALR(r)  \,, \pi(v) \big\}\Big\} \\ 
= & \ \min\Big\{\max\big\{bn(\ML_l)  \,, bn(\MR_r)  \big\}  \,, \max\big\{\infty  \,, \infty \,, |pq| \big\} \Big\}  \\
 = & \ \max\big\{bn(\ML_l)  \,, bn(\MR_r) \big\} = bn(\MLR_v) \,.
\end{align*}

%%%%%%%%%%%%%%%%%%%%%%%%%%%%%%%%%%%%%%%%%%%%%%%%%%%%%%%%%%%%%%%%%%%
%%  Case 3 $P_v$ is odd, $|P_l|$ is even, and $|P_r|$ is odd
%%%%%%%%%%%%%%%%%%%%%%%%%%%%%%%%%%%%%%%%%%%%%%%%%%%%%%%%%%%%%%%%%%%%
\noindent
\textbf{Case~3: \boldmath{$P_v$} is odd, $|P_l|$ is even, and $|P_r|$ is odd.} \\
Since $P_v$ is odd, there is no optimal matching $M_v$ of $P_v$, and thus $bn(M_v)=\infty$.
Moreover, $\MR_v$ consists of the optimal matching $M_l$ of $P_l$ and the optimal matching $\MR_r$ of $P_r\setminus\{\RM(r)\}$, and $\ML_v$ consists of the optimal matching $\MLR_l$ of $P_l\setminus\{\LM(l),\RM(l)\}$, the optimal matching $\ML_r$ of $P_r\setminus\{\LM(r)\}$, and the edge $(p,q)$.
Thus, $bn(\MR_v) = \max\{bn(M_l),bn(\MR_r)\}$ and $bn(\ML_v) = \max\{bn(\MLR_l),bn(\ML_r),|pq|\}$.

By the induction hypothesis, $\A(r)=\ALR(r)=\AL(l)=\AR(l)=\infty$, $\A(l)=bn(M_l)$, $\ALR(l)=bn(\MLR_l)$, $\AR(r)=bn(\MR_r)$, and $\AL(r)=bn(\ML_r)$. 
%Moreover, we need to show that $\AR(v)$ and $\AL(v)$ are $\infty$. 
Therefore, we have
\begin{align*}
 \A(r) = & \ \min\Big\{ \max\big\{\A(l) \,, \A(r)) \big\}  \,, \max\big\{\AR(l) \,, \AL(r)  \,, \pi(v) \big\}\Big\} \\
= & \ \min\Big\{ \max\big\{bn(M_l) \,, \infty \big\}  \,, \max\big\{\infty \,, bn(\ML_r) \,, |pq| \big\}\Big\} = \infty \,. \\
\AL(v) = & \ \min\Big\{ \max\big\{\AL(l) \,, \A(r) \big\}  \,, \\
& \ \quad \ \quad \ \max\big\{\ALR(l) \,, \AL(r)  \,, \pi(v) \big\}\Big\} \\
= & \ \min\Big\{ \max\big\{\infty \,, \infty \big\}  \,, \max\big\{bn(\MLR_l)  \,, bn(\ML_r) \,, |pq| \big\}\Big\} \\
= & \ \max\big\{bn(\MLR_l)  \,, bn(\ML_r) \,, |pq| \big\} = bn(\ML_v) \,. \\
\AR(v) = & \ \min\Big\{ \max\big\{\A(l) \,, \AR(r) \big\}  \,, \\
& \ \quad \ \quad \ \max\big\{\AR(l) \,, \ALR(r)  \,, \pi(v) \big\}\Big\} \\
= & \ \min\Big\{ \max\big\{bn(M_l) \,, bn(\MR_r) \big\}  \,, \max\big\{\infty \,, \infty  \,, |pq| \big\}\Big\}  \\
= & \ \max\big\{bn(M_l) \,, bn(\MR_r) \big\} = bn(\MR_v) \,. \\
\ALR(v) = & \ \min\Big\{ \max\big\{\AL(l) \,, \AR(r) \big\}  \,, \\
& \ \quad \ \quad \ \max\big\{\ALR(l) \,, \ALR(r)  \,, \pi(v) \big\}\Big\} \\ 
= & \ \min\Big\{\max\big\{\infty  \,, bn(\MR_r)  \big\}  \,, \max\big\{bn(\MLR_l)  \,, \infty \,, |pq| \big\} \Big\}  = \infty \,.
\end{align*}

%%%%%%%%%%%%%%%%%%%%%%%%%%%%%%%%%%%%%%%%%%%%%%%%%%%%%%%%%%%%%%%%%%%
%%  Case 4 $P_v$ is odd, $|P_l|$ is odd, and $|P_r|$ is even
%%%%%%%%%%%%%%%%%%%%%%%%%%%%%%%%%%%%%%%%%%%%%%%%%%%%%%%%%%%%%%%%%%%%
\noindent
\textbf{Case~4: \boldmath{$P_v$} is odd, $|P_l|$ is odd, and $|P_r|$ is even.} \\
This case is symmetric to Case~3.

\old{
%%%%%%%%%%%%%%%%%%%%%%%%%%%%%%%%%%%%%%%%%%%%%%%%%%%%%%%%%%%%%%%%%%%
%%  Case 2 $P_v$ is even and both $|P_l|$ and $|P_r|$ are odd
%%%%%%%%%%%%%%%%%%%%%%%%%%%%%%%%%%%%%%%%%%%%%%%%%%%%%%%%%%%%%%%%%%%%
\noindent
\textbf{Case~2: \boldmath{$|P_v|$} is even and both $|P_l|$ and $|P_r|$ are odd.} \\
Since $|P_v|$ is even, $M_v$ consists of the optimal matching $\ML_r$ of $P_r\setminus\{\LM(r)\}$, the optimal matching $\MR_l$ of $P_l\setminus\{\RM(l)\}$, and the edge $(p,q)$. Thus, $bn(M_v) = \max\{bn(\ML_r),bn(\MR_l),|pq|\}$.
Moreover, $\MLR_v$ consists of the optimal matching $\ML_l$ of $P_l\setminus\{\LM(l)\}$ and the optimal matching $\MR_r$ of $P_r\setminus\{\RM(r)\}$, and $bn(\MLR_v) = \max\{bn(\ML_l),bn(\MR_r)\}$.

By the induction hypothesis, $\A(l)=\ALR(l)=\A(r)=\ALR(r)=\infty$, $\AR(l)=bn(\MR_l)$, $\AL(l)=bn(\ML_l)$, $\AR(r)=bn(\MR_r)$, and $\AL(r)=bn(\ML_r)$. 
%Moreover, we need to show that $\AR(v)$ and $\AL(v)$ are $\infty$. 
Therefore, we have
\begin{align*}
 \A(r) = & \ \min\Big\{ \max\big\{\A(l) \,, \A(r)) \big\}  \,, \max\big\{\AR(l) \,, \AL(r)  \,, \pi(v) \big\}\Big\} \\
= & \ \min\Big\{ \max\big\{\infty \,, \infty \big\}  \,, \max\big\{bn(\MR_l) \,, bn(\ML_r)  \,, |pq| \big\}\Big\}  \\
= & \ \max\big\{bn(\MR_l) \,,bn(\ML_r) \,, |pq| \big\} = bn(M_v) \,. \\
\AL(v) = & \ \min\Big\{ \max\big\{\AL(l) \,, \A(r) \big\}  \,, \\
& \ \quad \ \quad \ \max\big\{\ALR(l) \,, \AL(r)  \,, \pi(v) \big\}\Big\} \\
= & \ \min\Big\{ \max\big\{bn(\ML_l) \,, \infty \big\}  \,, \max\big\{\infty  \,, bn(\ML_r) \,, |pq| \big\}\Big\} = \infty \,. \\
\AR(v) = & \ \min\Big\{ \max\big\{\A(l) \,, \AR(r) \big\}  \,, \\
& \ \quad \ \quad \ \max\big\{\AR(l) \,, \ALR(r)  \,, \pi(v) \big\}\Big\} \\
= & \ \min\Big\{ \max\big\{\infty \,, bn(\MR_r) \big\}  \,, \max\big\{bn(\MR_l) \,, \infty  \,, |pq| \big\}\Big\} = \infty \,. \\
\ALR(v) = & \ \min\Big\{ \max\big\{\AL(l) \,, \AR(r) \big\}  \,, \\
& \ \quad \ \quad \ \max\big\{\ALR(l) \,, \ALR(r)  \,, \pi(v) \big\}\Big\} \\ 
= & \ \min\Big\{\max\big\{bn(\ML_l)  \,, bn(\MR_r)  \big\}  \,, \max\big\{\infty  \,, \infty \,, |pq| \big\} \Big\}  \\
 = & \ \max\big\{bn(\ML_l)  \,, bn(\MR_r) \big\} = bn(\MLR_v) \,.
\end{align*}

%%%%%%%%%%%%%%%%%%%%%%%%%%%%%%%%%%%%%%%%%%%%%%%%%%%%%%%%%%%%%%%%%%%
%%  Case 3 $P_v$ is odd, $|P_l|$ is even, and $|P_r|$ is odd
%%%%%%%%%%%%%%%%%%%%%%%%%%%%%%%%%%%%%%%%%%%%%%%%%%%%%%%%%%%%%%%%%%%%
\noindent
\textbf{Case~3: \boldmath{$P_v$} is odd, $|P_l|$ is even, and $|P_r|$ is odd.} \\
Since $P_v$ is odd, there is no optimal matching $M_v$ of $P_v$, and thus $bn(M_v)=\infty$.
Moreover, $\MR_v$ consists of the optimal matching $M_l$ of $P_l$ and the optimal matching $\MR_r$ of $P_r\setminus\{\RM(r)\}$, and $\ML_v$ consists of the optimal matching $\MLR_l$ of $P_l\setminus\{\LM(l),\RM(l)\}$, the optimal matching $\ML_r$ of $P_r\setminus\{\LM(r)\}$, and the edge $(p,q)$.
Thus, $bn(\MR_v) = \max\{bn(M_l),bn(\MR_r)\}$ and $bn(\ML_v) = \max\{bn(\MLR_l),bn(\ML_r),|pq|\}$.

By the induction hypothesis, $\A(r)=\ALR(r)=\AL(l)=\AR(l)=\infty$, $\A(l)=bn(M_l)$, $\ALR(l)=bn(\MLR_l)$, $\AR(r)=bn(\MR_r)$, and $\AL(r)=bn(\ML_r)$. 
%Moreover, we need to show that $\AR(v)$ and $\AL(v)$ are $\infty$. 
Therefore, we have
\begin{align*}
 \A(r) = & \ \min\Big\{ \max\big\{\A(l) \,, \A(r)) \big\}  \,, \max\big\{\AR(l) \,, \AL(r)  \,, \pi(v) \big\}\Big\} \\
= & \ \min\Big\{ \max\big\{bn(M_l) \,, \infty \big\}  \,, \max\big\{\infty \,, bn(\ML_r) \,, |pq| \big\}\Big\} = \infty \,. \\
\AL(v) = & \ \min\Big\{ \max\big\{\AL(l) \,, \A(r) \big\}  \,, \\
& \ \quad \ \quad \ \max\big\{\ALR(l) \,, \AL(r)  \,, \pi(v) \big\}\Big\} \\
= & \ \min\Big\{ \max\big\{\infty \,, \infty \big\}  \,, \max\big\{bn(\MLR_l)  \,, bn(\ML_r) \,, |pq| \big\}\Big\} \\
= & \ \max\big\{bn(\MLR_l)  \,, bn(\ML_r) \,, |pq| \big\} = bn(\ML_v) \,. \\
\AR(v) = & \ \min\Big\{ \max\big\{\A(l) \,, \AR(r) \big\}  \,, \\
& \ \quad \ \quad \ \max\big\{\AR(l) \,, \ALR(r)  \,, \pi(v) \big\}\Big\} \\
= & \ \min\Big\{ \max\big\{bn(M_l) \,, bn(\MR_r) \big\}  \,, \max\big\{\infty \,, \infty  \,, |pq| \big\}\Big\}  \\
= & \ \max\big\{bn(M_l) \,, bn(\MR_r) \big\} = bn(\MR_v) \,. \\
\ALR(v) = & \ \min\Big\{ \max\big\{\AL(l) \,, \AR(r) \big\}  \,, \\
& \ \quad \ \quad \ \max\big\{\ALR(l) \,, \ALR(r)  \,, \pi(v) \big\}\Big\} \\ 
= & \ \min\Big\{\max\big\{\infty  \,, bn(\MR_r)  \big\}  \,, \max\big\{bn(\MLR_l)  \,, \infty \,, |pq| \big\} \Big\}  = \infty \,.
\end{align*}

%%%%%%%%%%%%%%%%%%%%%%%%%%%%%%%%%%%%%%%%%%%%%%%%%%%%%%%%%%%%%%%%%%%
%%  Case 4 $P_v$ is odd, $|P_l|$ is odd, and $|P_r|$ is even
%%%%%%%%%%%%%%%%%%%%%%%%%%%%%%%%%%%%%%%%%%%%%%%%%%%%%%%%%%%%%%%%%%%%
\noindent
\textbf{Case~4: \boldmath{$P_v$} is odd, $|P_l|$ is odd, and $|P_r|$ is even.} \\
This case is symmetric to Case~3.
}
\end{proof}

%%%%%%%%%%%%%%%%%%%%%%%%%%%%%%%%%%%%%%%%%%%%%%%%%%%%%%%%%%%%%%%%%%%%
%\subsection{The Data  structure}\label{sec:3-1}
\subsection{Dynamization} \label{sec:2-2}
Let $P=\{p_1,p_2,\ldots,p_n\}$ be the set of points at some time step and let $\T$ be the AVL tree maintaining the optimal matching $M$ of $P$. Let $r$ denote the root of $\T$. 
In the following, we describe how to update $\T$ when inserting two points to $P$ or deleting two points from $P$.

\subsection*{Insertion}
\vspace{-0.1cm}
Let $q$ and $q'$ be the two points inserted to $P$.
We describe the procedure for inserting $q$. The same procedure is applied for inserting $q'$.
We initialize a leaf node corresponding to $q$ and insert it to $\T$.
Then, we update the attributes of the intermediate nodes along the path from $q$ to the root of $\T$.

Let $M'$ be the optimal matching of $P\cup\{q,q'\}$. Then, by Lemma~\ref{lemma:1Dcorrect}, after inserting $q$ and $q'$ to $P$, $\A(r) = bn(M')$.

\subsection*{Deletion}
\begin{figure}[htb]
    \centering
    \includegraphics[width=0.8\textwidth]{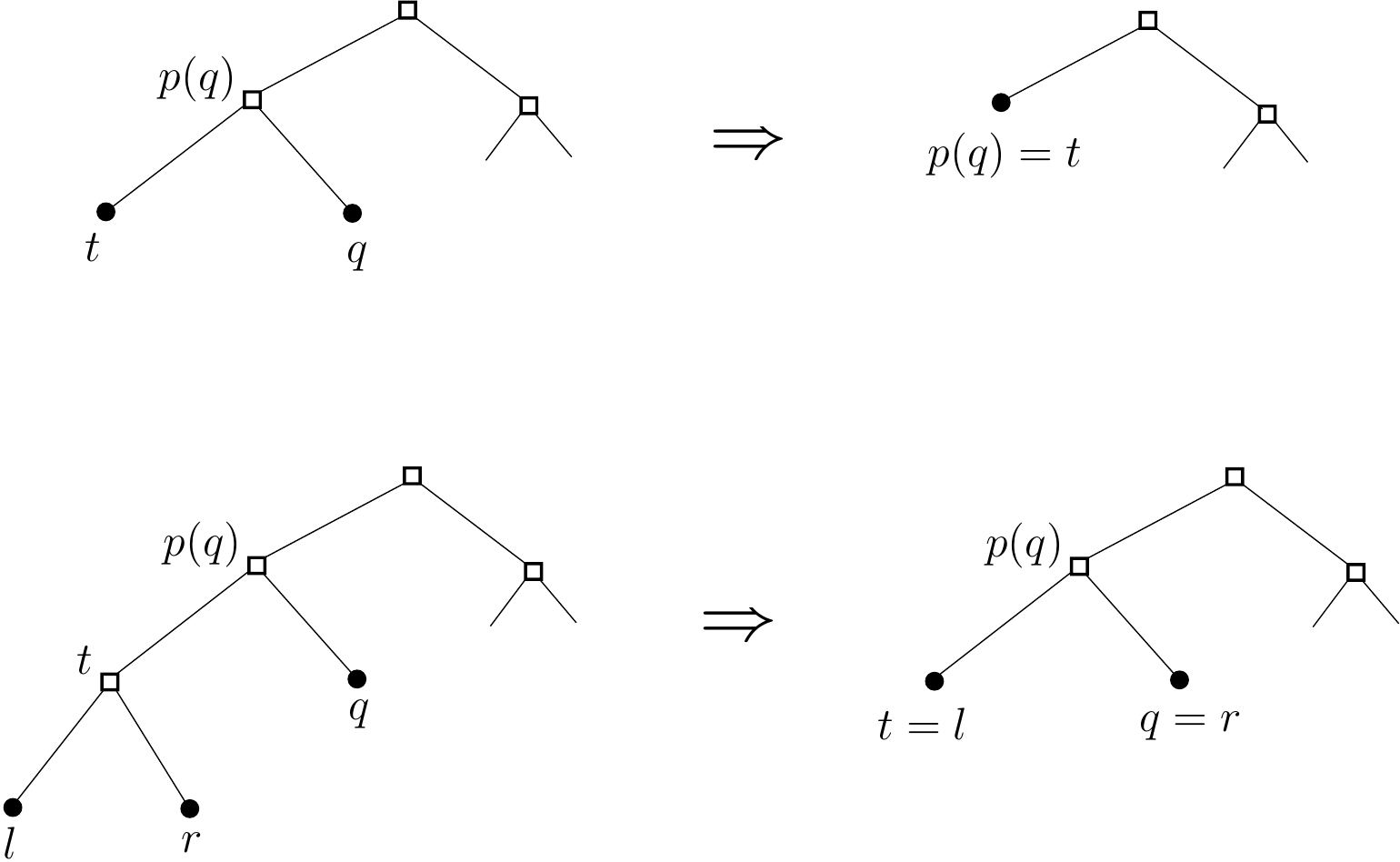}
    \caption{Deleting $q$ from $\T$.}
    \label{fig:deletion}
\end{figure}

\vspace{-0.1cm}
Let $q$ and $q'$ be the two points deleted from $P$.
We describe the procedure for deleting $q$. The same procedure is applied for deleting $q'$.
Assume w.l.o.g. that $q$ is the right child of $p(q)$.
If the left child $t$ of $p(q)$ is a leaf, then we set the attributes of $t$ to $p(q)$, remove $q$ and $t$ from $\T$, and update the attributes of the intermediate nodes along the path from $p(q)$ to the root of $\T$; see Figure~\ref{fig:deletion}(top).
Otherwise, the left child $t$ of $p(q)$ is an intermediate node with left leaf $l$ and right leaf $r$.
We set the attributes of $l$ to $t$ and the attributes of $r$ to $q$, remove $l$ and $r$ from $\T$, and update the attributes of the intermediate nodes along the path from $p(q)$ to the root of $\T$; see Figure~\ref{fig:deletion}(bottom).

Let $M'$ be the optimal matching of $P\setminus\{q,q'\}$. Then, by Lemma~\ref{lemma:1Dcorrect}, after deleting $q$ and $q'$ from $P$, $\A(r) = bn(M')$.

Finally, since we use an AVL tree, we may need to make some rotations after an insertion or a deletion.
For each rotation performed on $\T$, we also update the attributes of the (constant number of) intermediate nodes involved in the rotation. 

\begin{lemma}
The running time of an update operation (insertion or deletion) is $O(\log{n})$.
\end{lemma}
\begin{proof}
Since $\T$ is an AVL tree, the height of $\T$ is $O(\log{n})$~\cite{Cormen09}.
Each operation requires updating the attributes of the nodes along the path from a leaf to the root, and each such update takes $O(1)$ time. Moreover, each rotation also requires updating the attributes of the nodes involved in the rotation, and each such update also takes $O(1)$ time .
Since in insertion there is at most one rotation and in deletion there are at most $O(\log{n})$ rotations, the total running time 
of each insertion and each deletion is $O(\log{n})$.
\end{proof}
 
The following theorem summarizes the result of this section.
\begin{theorem}
Let $P$ be a set of $n$ points on a line. There exists a dynamic algorithm that maintains a bottleneck matching of $P$ and supports insertion and deletion in $O(\log n)$ time.
\end{theorem}

%%%%%%%%%%%%%%%%%%%%%%%%%%%%%%%%%%%%%%%%%%%%%%%%%%%%%%%%%%%%%%%%%%%%%%%%%%%%%%%%%%%%%%%5%%%%%%%%%%

\section{Extensions for 1D}\label{sec:Ext}
In this section, we extend our algorithm to maintain a minimum-weight matching of $P$ (instead of bottleneck matching). Moreover, we extend the algorithm to allow inserting/deleting a constant (even or odd) number of points to/from $P$.

\subsection{Minimum-weight matching}
%\vspace{-0.1cm}
We modify our algorithm to maintain a minimum-weight matching and support insertion and deletion, without affecting the running time.
%The algorithm we described to maintain the bottleneck matching that supports insertion and deletion can be generalized  to maintain minimum-weight matching that supports insertion and deletion, without affecting the running time.
The difference lies in the way we compute the attributes of the intermediate nodes from their children. 
That is, for each intermediate node $v$, we compute its attributes as follows:
%Now, we describe how to compute the entries of $V_{v}[4]$ in each node $v$, from the entries of its children $lc(t)$ and $rc(t)$. 
%
\begin{align*}
\A(v) \leftarrow &  \min\big\{\A(lc(v)) + \A(rc(v))   \,, \\
& \quad \ \quad \ \AR(lc(v)) + \AL(rc(v))  +  \pi(v) \big\}\,.\\ 
\AL(v) \leftarrow &  \min\big\{\AL(lc(v)) + \A(rc(v))   \,, \\
& \quad \ \quad \ \ALR(lc(v)) + \AL(rc(v))  + \pi(v) \big\}\,.
\end{align*}
\begin{align*}
\AR(v) \leftarrow &  \min\big\{ \A(lc(v)) + \AR(rc(v))   \,, \\
& \quad \ \quad \ \AR(lc(v)) + \ALR(rc(v))  + \pi(v) \big\}\,.\\
\ALR(v) \leftarrow &  \min\big\{ \AL(lc(v)) + \AR(rc(v))   \,, \\
& \quad \ \quad \ \ALR(lc(v)) + \ALR(rc(v)) + \pi(v) \big\}\,. 
\end{align*}

Notice that the running time of an update operation ($O(\log{n})$ per insertion or deletion) is as 
in the bottleneck matching. The proof of the correctness of this algorithm for the minimum-weight matching is similar to the proof of the correctness of the bottleneck matching.  

\subsection{Insertion and deletion of $k$ points}
%\vspace{-0.1cm}
Let $P$ be a set of $n$ points on a line.
In this section, we extend our algorithm to support insertion/deletion of $k$ points to/from $P$ at each time step.
%The algorithm we described to maintain the bottleneck matching that supports insertion and deletion of two points can be generalized to maintain insertion and deletion of $k$ points, such that its update time is $O(k \log{n})$. 
 Notice that since we allow $k$ to be odd, $n$ can be odd and the matching should skip one point. Even though there are linear different candidate points that could be skipped, we can still maintain a bottleneck matching with $O(k \log{n})$ time per $k$ insertions or deletions, by adding some more attributes for each node. 
Each node $v$ in $\T$ contains the following four attributes, in addition to the seven attributes that are described in Section~\ref{sec:1DPreprocessing}.
 \begin{enumerate}
    %\vspace{-.05cm}
    \item[8.] $\Miss(v)$ - cost of the matching of $|P_v|-1$ points of $P_v$.
    %\vspace{.05cm}
    \item[9.] $\MissL(v)$ - cost of the matching of $|P_v|-1$ points of $P_v\setminus\{\LM(v)\}$.
    %\vspace{.05cm}
    \item[10.] $\MissR(v)$ - cost of the matching of $|P_v|-1$ points of $P_v\setminus\{\RM(v)\}$.
    %\vspace{.05cm}
    \item[11.] $\MissLR(v)$ - cost of the matching of $|P_v|-1$ points of $P_v\setminus\{\LM(v),$\\$\RM(v)\}$.
\end{enumerate}
 
 For each leaf $v$ in $\T$, we initialize $\Miss(v)$ to be 0, and $\MissL(v)$, $\MissR(v)$, and $\MissLR(v)$ to be $\infty$.
 For each intermediate node $v$ in $\T$, we compute its attributes as follows.

\begin{align*}
\A(v) \leftarrow &  \min\Big\{ \max\big\{\A(lc(v)) \,, \A(rc(v)) \big\}  \,, \\
& \quad \ \quad \ \max\big\{\AR(lc(v)) \,, \AL(rc(v))  \,, \pi(v) \big\}\Big\}\,.\\ 
\AL(v) \leftarrow &  \min\Big\{ \max\big\{\AL(lc(v)) \,, \A(rc(v)) \big\}  \,, \\
& \quad \ \quad \ \max\big\{\ALR(lc(v)) \,, \AL(rc(v))  \,, \pi(v) \big\}\Big\}\,.\\
\AR(v) \leftarrow &  \min\Big\{ \max\big\{\A(lc(v)) \,, \AR(rc(v)) \big\}  \,, \\
& \quad \ \quad \ \max\big\{\AR(lc(v)) \,, \ALR(rc(v))  \,, \pi(v) \big\}\Big\}\,.\\
\ALR(v) \leftarrow &  \min\Big\{ \max\big\{\AL(lc(v)) \,, \AR(rc(v)) \big\}  \,, \\
& \quad \ \quad \ \max\big\{\ALR(lc(v)) \,, \ALR(rc(v))  \,, \pi(v) \big\}\Big\}\,. \\ 
\Miss(v) \leftarrow &  \min\Big\{ \max\big\{\Miss(lc(v)) \,, \A(rc(v)) \big\}  \,, \\
& \quad \ \quad \ \max\big\{\A(lc(v)) \,, \Miss(rc(v)) \big\} \,,\\ 
& \quad \ \quad \ \max\big\{\MissR(lc(v)) \,, \AL(rc(v))  \,, \pi(v) \big\} \,,\\
& \quad \ \quad \ \max\big\{\AR(lc(v)) \,, \MissL(rc(v))  \,, \pi(v) \big\}\Big\} \,.\\
\MissL(v) \leftarrow &  \min\Big\{ \max\big\{\MissL(lc(v)) \,, \A(rc(v)) \big\}  \,, \\
& \quad \ \quad \ \max\big\{\AL(lc(v)) \,, \Miss(rc(v))  \big\} \,,\\ 
& \quad \ \quad \ \max\big\{\MissLR(lc(v)) \,, \AL(rc(v))  \,, \pi(v) \big\} \,,\\ 
& \quad \ \quad \ \max\big\{\ALR(lc(v)) \,, \MissL(rc(v))  \,, \pi(v) \big\}\Big\}\,.\\ 
\MissR(v) \leftarrow &  \min\Big\{ \max\big\{\Miss(lc(v)) \,, \AR(rc(v)) \big\}  \,, \\
& \quad \ \quad \ \max\big\{\A(lc(v)) \,, \MissR(rc(v))  \big\} \,,\\ 
& \quad \ \quad \ \max\big\{\MissR(lc(v)) \,, \ALR(rc(v))  \,, \pi(v) \big\} \,,\\ 
& \quad \ \quad \ \max\big\{\AR(lc(v)) \,, \MissLR(rc(v))  \,, \pi(v) \big\}\Big\}\,.\\ 
\end{align*}
\begin{align*}
\MissLR(v) \leftarrow &  \min\Big\{ \max\big\{\MissL(lc(v)) \,, \AR(rc(v)) \big\}  \,, \\
& \quad \ \quad \ \max\big\{\AL(lc(v)) \,, \MissR(rc(v)) \big\} \,, \\ 
& \quad \ \quad \ \max\big\{\MissLR(lc(v)) \,, \ALR(rc(v))  \,, \pi(v) \big\} \,, \\ 
& \quad \ \quad \ \max\big\{\ALR(lc(v)) \,, \MissLR(rc(v))  \,, \pi(v) \big\}\Big\}\,. 
\end{align*}

Let $r^*$ be the root of $\T$. 
In the case that $n$ is even, let $M$ be the bottleneck matching for $P_{r^*}$ 
satisfying Observation~\ref{obs:sec-1}. 
In the case that $n$ is odd, 
let $M_q$ be the bottleneck matching for $P_{r^*} \setminus \{q \}$ 
satisfying Observation~\ref{obs:sec-1}. 
Let $M'$ the  bottleneck matching such that $bn(M') =\min_{q \in P_{r^*}} \{bn(M_q) \} $.
%.
\begin{lemma} \label{lemma:1DcorrectEvenAndOdd}
Let $r^*$ be the root of $\T$. 
\begin{itemize}
    \item If $n$ is even, then $\A(r^*)=bn(M)$.
    \item If $n$ is odd, then $\Miss(r^*)=bn(M')$.
\end{itemize}
\end{lemma}

The proof of Lemma~\ref{lemma:1DcorrectEvenAndOdd} is similar to the proof of Lemma~\ref{lemma:1Dcorrect}.
Moreover, the insertion and the deletion operations are done as in Section~\ref{sec:2-2}. After each operation, we update the attributes (including the new attributes) of the intermediate nodes along the path from a leaf to the root.
%one needs to show that the insertion or deletion of one point  updates the attributes in each node correctly, and this proof is similar to the proof of the correctness bottleneck matching where $n$ is even.

The running time of $O(k \log{n})$ is obtained by performing the operation (insertion or deletion) $k$ times. That is, when we are requested to insert/delete $k$ points we add/remove them one by one. 
Thus, the $O(\log{n})$ time per update operation is performed $k$ times.

%%%%%%%%%%%%%%%%%%%%%%%%%%%%%%%%%%%%%%%%%%%%%%%%%%%%%%%%%%%%%%%%%%%%%%%%%%%%%%%%%%%%%%%%%%%%%%%%%%
\section{Dynamic Bottleneck Matching in 2D}
\label{Sec:Dynmic2D}
%In this section, we consider a set of dynamic points in the plane. 
Let $\P= P_1 \cup P_2 \cup \dots $ be a set of $n$ points in the plane, 
such that each set $P_{k+1}$ is obtained by adding a pair of points 
to $P_k$ or by removing a pair of points from $P_k$.
Let $\lambda_k$ be the distance between the closest pair of points in $P_k$.
In our setting, we assume that we are given a bounding box $\B$ of side length $\Lambda$ and a constant $\lambda > 0$, such that $P_k$ is contained in $\B$ and $\lambda \le \lambda_k$, for each $k \ge 1$, and $\Delta =  {\Lambda \over \lambda}$ is polynomially bounded in $n$, i.e., $\log{\Delta}=O(\log{n})$. 
\old{
Let $\P$ be a set of $n$ points in the plane.
Let $\lambda$ be the distance between the closest pair of points in $P$ and let $\Lambda$ be the distance between the farthest pair of points in $P$.  
Let $\Delta$ be the geometric spread of $P$, that is $\Delta =  {\Lambda \over \lambda}$.

We assume that $\lambda$ and $\Lambda$ are given and that $\Delta$ is polynomially bounded in $n$, i.e., $\log{\Delta}=O(\log{n})$.
}

%Moreover, we assume that $\Delta$ is is polynomially bounded in $n$, i.e., $\log{\Delta}=O(\log{n})$.
%, that is $\Delta_{P_k}\le \poly(n)$, $\forall k\in \mathbb{N}$.
At each time step~$k\in \mathbb{N}$, either a pair of points of $P$ is inserted or deleted.
Let $P_k$ be the set of points at time step~$k$ and let $M^{*}_k$ be a bottleneck matching of $P_k$ of bottleneck $bn(M^*_k)$.
In this section, we present a dynamic data structure supporting insertion in $O(\log{\Delta})$ time and deletion in $O(\log{\Delta} + |P_k|)$ time, such that a perfect matching $M_k$ of $P_k$ of bottleneck at most $6\sqrt{2}\cdot bn(M^*_k)$ can be computed in $O(\log{\Delta} + |P_k|)$ time.

\old{
We first describe (in Section~\ref{sec:3-1}) the data structure 
that we construct and maintain in order to update efficiently and answer 
 in linear time an approximate bottleneck matching $M_k$ of $P_k$ in the static case. Then, in Section~\ref{sec:3-2}, we describe the dynamic data structure supporting insertion and deletion operations at each step.
}

Let $\B$ be the bounding box containing the points of $\P$.
%, and let $\lambda_{P}$ be the distance of the closest pair of points in $P$. 
Set $c=\ceil{\log \Delta}$.
For each integer $0\le i \le c$, let $\Pi_i$ be the grid obtained by dividing $\B$ into cells of side length $2^{i}\cdot \lambda$.
We say that two cells are adjacent in $\Pi_i$ if they share a side or a corner in $\Pi_i$.

Let $P=P_k$ be the set of points at some time step~$k$. %and let $M^{*}$ be a bottleneck matching of $P$ of bottleneck $bn(M^*)$.
For each grid $\Pi_i$, we define an undirected graph $G_i$, such that the vertices of $G_i$ are the non-empty cells of $\Pi_i$, and there is an edge between two non-empty cells in $G_i$ if these cells are adjacent in $\Pi_i$. 
%For $S\subseteq P$, let $C(S,\Pi_i)$ be the cells of $\Pi_i$ 
For a vertex $v$ in $G_i$, let $P_v$ be the set of points of $P$ that are contained in the cell in $\Pi_i$ corresponding to $v$.
For a connected component $C$ in $G_i$, let $P(C)=\bigcup_{v\in C} P_v$, i.e., the set of the points contained in the cells corresponding to the vertices of $C$.
Moreover, we assume that each graph $G_i$ has a parity bit that indicates whether all the connected components of $G_i$ contain an even number of points or not.

\begin{lemma}\label{lemma:upperbound}
Let $C$ be a connected component in $G_i$. If $|P(C)|$ is even, then there exists a perfect matching of the points of $P(C)$ of bottleneck at most $3\sqrt{2}\cdot 2^{i}\cdot \lambda$. Moreover, this matching can be computed in $O(|P(C)|)$ time. 
\end{lemma}
\begin{proof}
Let $G_C$ be the subgraph of $G_i$ induced by $C$.
Let $T$ be a spanning tree of $G_C$ and assume that $T$ is rooted at a vertex $r$.
We construct a perfect matching of the points of $P(C)$ iteratively by considering $T$ bottom-up as follows.
Let $v$ be the deepest vertex in $T$ which is not a leaf, and let $v_1,v_2,\dots,v_j$ be its children in $T$. Notice that $v_1,v_2,\dots,v_j$ are leaves.
Let $P'= \bigcup_{1\le i \le j} P_{v_i}$ be the set of the points contained in the cells corresponding to $v_1,v_2,\dots,v_j$. 
If $|P'|$ is even, then we greedily match the points in $P'$ and remove the vertices $v_1,v_2,\dots,v_j$ from $T$.
Otherwise, $|P'|$ is odd. In this case, we select an arbitrary point $p$ from the cell corresponding to $v$ and greedily match the points in $P'\cup \{p\}$.
Moreover, we remove $p$ from the cell corresponding to $v$ and remove $v_1,v_2,\dots,v_j$ from $T$. 
We continue this procedure until the root $r$ is encountered, i.e., until $v=r$.

Since $|P(C)|$ is even and in each iteration, we match an even number of points, the number of the points in the last iteration is even and we get a perfect matching of the points of $P(C)$. 
Moreover, since in each iteration we match points from the cell corresponding to $v$ and its at most eight neighbors in $\Pi_i$, and these cells are contained in $3 \times 3$ cells-block, the length of each edge in the matching is at most $3\sqrt{2}\cdot 2^{i}\cdot \lambda$.

Since the degree of each vertex in $G_C$ is at most eight, computing $T$ takes $O(|C|)$, and matching the points of $P'$ in each iteration takes $O(|P'|)$. Therefore, computing the matching of the points of $P(C)$ takes $O(|P(C)|)$ time.
\end{proof}

Let $M^*$ be a bottleneck matching of $P$ and let $bn(M^*)$ be its bottleneck.

\begin{lemma}\label{lemma:lowerbound}
If $bn(M^*) \le 2^{i}\cdot \lambda$, then, for every connected component $C$ in $G_i$, $|P(C)|$ is even.
\end{lemma}
\begin{proof}
Assume by contradiction that there is a connected component $C$ in $G_i$, such that $|P(C)|$ is odd. Thus, at least one point $p \in P(C)$ is matched in $M^*$ to a point $q\notin P(C)$. Therefore, $|pq| > 2^{i}\cdot \lambda$, which contradicts that $bn(M^*) \le 2^{i}\cdot \lambda$.
\end{proof}

%By Lemma~\ref{lemma:upperbound} and Lemma~\ref{lemma:lowerbound}, we have the following theorem. 
%
\begin{theorem}
In $O(\log{\Delta})$ time we can compute a value $t$, such that $t < bn(M^*) \le 6\sqrt{2}\cdot t$.
Moreover, we can compute a perfect matching $M$ of $P$ of bottleneck at most $6\sqrt{2}\cdot bn(M^*)$ in $O(\log{\Delta} + |P|)$ time.
\end{theorem}
\begin{proof}
Let $i$ be the smallest integer such that all the connected components in $G_i$ have an even number of points. 
Thus, by Lemma~\ref{lemma:lowerbound}, $bn(M^*) > 2^{i-1}\cdot \lambda$, and, by Lemma~\ref{lemma:upperbound}, there exists a perfect matching of $P$ of bottleneck at most $3\sqrt{2}\cdot 2^{i}\cdot \lambda$.
Therefore, by taking $t = 2^{i-1}\cdot \lambda$, we have $t < bn(M^*) \le 6\sqrt{2}\cdot t$.
Since each graph $G_i$ has a parity bit, we can compute $t$ in $O(\log{\Delta})$ time.
Moreover, by Lemma~\ref{lemma:upperbound}, we can compute a perfect matching $M$ of $P$ of bottleneck at most $3\sqrt{2}\cdot 2^{i}\cdot \lambda$ in $O(|P|)$ time. Therefore, $bn(M) \le 3\sqrt{2}\cdot 2^{i}\cdot \lambda \le 6\sqrt{2}\cdot bn(M^*)$. 
\end{proof}

%%%%%%%%%%%%%%%%%%%%%%%%%%%%%%%%%%%%%%%%%%%%%%%%%%%%%%%%%%%%%%%%%%%%
\subsection{Preprocessing}
We first introduce a data structure that will be used in the preprocessing.
\subsection*{Disjoint-set data structure} 
\vspace{-0.1cm}
A \emph{disjoint-set data structure} is a data structure that maintains a collection $D$ of disjoint
dynamic sets of objects and each set in $D$ has a representative, which is some member of the set (see~\cite{Cormen09} for more details).
Disjoint-set data structures support the following operations:
\begin{itemize}
    \item \textsc{Make-Set}$(x)$ creates a new set whose only member (and thus representative) is the object $x$.
    \item \textsc{Union}$(S_i,S_j)$ merges the sets $S_i$ and $S_j$ and choose either the representative of $S_i$ or the representative of $S_j$ to be the representative of the resulting set.
    \item \textsc{Find-Set}$(x)$ returns the representative of the (unique) set containing $x$.
\end{itemize}

It has been proven in ~\cite{Tarjan84} that performing a sequence of $m$ \textsc{Make-Set}, \textsc{Union}, or \textsc{Find-Set} operations on a disjoint-set data structures with $n$ objects requires total time $O(m\cdot\alpha(n))$, where $\alpha(n)$ is the extremely slow-growing \emph{inverse Ackermann function}.
More precisely, it has been shown that the amortized time of each one of the operations \textsc{Make-Set}, \textsc{Union}, and \textsc{Find-Set} is $O(1)$. 

We associate each set $S$ in $D$ with a variable $v_S$ that represents the parity of $S$ depending on the number of points in $S$. 
We also modify the operations \textsc{Make-Set}$(x)$ to initialize the parity variable of the created set to be odd, and \textsc{Union}$(S_i,S_j)$ to update the parity variable of the joined set according to the parities of $S_i$ and $S_j$. 
Moreover, we define a new operation \textsc{Change-Parity}$(S)$ that inverses the parity of the set $S$. Notice that these
changes do not affect the performance of the data structure.

We now describe how to initialize our data structure, given the bounding box $\B$, the constant $\lambda$, and an initial set $P_1$. 
Set $c=\ceil{\log \Delta}$.
For each integer $0\le i \le c$, let $\Pi_i$ be the grid obtained by dividing $\B$ into cells of side length $2^{i}\cdot \lambda$.
For each grid $\Pi_i$, we use a disjoint-set data structure $DSS_i$ to maintain the connected components of $G_i$ that is defined on $\Pi_i$ and $P_1$.
That is, the objects of $DSS_i$ are the non-empty cells of $\Pi_i$, and if two non-empty cells share a side or a corner in $\Pi_i$, then they are in the same set in $DSS_i$.
This data structure guarantees that each connected component in $G_i$ is a set in $DSS_i$.
%the objects of $DSS_i$ are the vertices of $G_i$ (i.e., the non-empty cells of $\Pi_i$) and the sets of $DSS_i$ are the connected components of $G_i$.

%if two non-empty cells share a side or a corner in $\Pi_i$, then they are in the same set in $DSS_i$. This data structure guarantees that each connected component in $G_i$ (the cells graph defined on $\Pi_i$) is a set in $DSS_i$.

As mentioned above, constructing each $DSS_i$ can be done in $O(|P_1|)$ time. Therefore, the preprocessing time is $O(\log{\Delta}\cdot |P_1|)$.
% can be improved to $O(\log{\Delta} + \log{\log{\Delta}}\cdot |P_1|)$??

%%%%%%%%%%%%%%%%%%%%%%%%%%%%%%%%%%%%%%%%%%%%%%%%%%%%%%%%%%%%%%%%%%%%
%\subsection{The Data  structure}\label{sec:3-1}
\subsection{Dynamization} \label{sec:3-2}
Let $P$ be the set of points at some time step. 
In the following, we describe how to update each structure $DSS_i$ when inserting two points to $P$ or deleting two points from $P$.  

%\vspace{-0.3cm}
\subsection*{Insertion} 
\vspace{-0.1cm}
Let $p$ and $q$ be the two points inserted to set $P$.
We describe the procedure for inserting $p$. The same procedure is applied for inserting $q$.
For each grid $\Pi_i$, we do the following; see Procedure~\ref{alg:insert}.
Let $Cell_{i}(p)$ be the cell containing $p$ in $\Pi_i$.
If $Cell_{i}(p)$ contains points of $P$, then we find the set containing $Cell_{i}(p)$ in $DSS_i$ and change its parity.
Otherwise, we make a new set in $DSS_i$ containing the cell $Cell_{i}(p)$ and merge (union) it with all the sets in $DSS_i$ that contain a non-empty adjacent cell of $Cell_{i}(p)$, and update the parity of the joined set. 

\vspace{-0.1cm}
\floatname{algorithm}{Procedure}
\begin{algorithm}[htb]
\caption{\textsc{Insert($p$)}} \label{alg:insert}

\begin{algorithmic}[1]

\STATE \textbf{for each} $0 \le i \le c$ \textbf{do} \\ 
    \STATE \quad \ $Cell_{i}(p) \leftarrow$ the cell containing $p$ in $\Pi_i$  \\
    \STATE \quad \ \textbf{if} $Cell_{i}(p) \cap P = \emptyset$ \textbf{then} \hfill /* $Cell_{i}(p)$ contains only $p$ */\\
        \STATE \quad \ \quad \ \textsc{Make-Set($Cell_{i}(p)$)}  \\
        \STATE \quad \ \quad \ \textbf{for each} non-empty adjacent cell $C$ of $Cell_{i}(p)$ \textbf{do} \\
            \STATE \quad \ \quad \ \quad \ $S_C \leftarrow$ \textsc{Find-Set($C$)} \\
            \STATE \quad \ \quad \ \quad \ $S_p \leftarrow$ \textsc{Find-Set($Cell_{i}(p)$)} \\
            \STATE \quad \ \quad \ \quad \ \textsc{Union($S_C,S_p$)} \\
    \STATE \quad \ \textbf{else} \hfill /* $Cell_{i}(p)$ contains points other than $p$ */ \\
        \STATE \quad \ \quad \ $S_i(p) \leftarrow$ \textsc{Find-Set($Cell_{i}(p)$)} \\
        \STATE \quad \ \quad \ \textsc{Change-Parity($S_i(p)$)} 

\end{algorithmic}
\end{algorithm}

%\vspace{-0.3cm}
\begin{lemma}
\textsc{Insert($p$)} takes amortized $O(\log{\Delta})$ time.
\end{lemma}
\begin{proof}
Finding the cell containing $p$ in each grid $\Pi_i$ can be done in constant time.
If $Cell_{i}(p)$ contains points of $P$, then we change the parity of the set containing $Cell_{i}(p)$ in $DSS_i$ in constant time.
Otherwise, making a new set in $DSS_i$ and merging it with at most eight sets in $DSS_i$ that contain non-empty adjacent cells of $Cell_{i}(p)$ can be also done in amortized constant time.
Since $c=\ceil{\log \Delta}$, \textsc{Insert($p$)} takes amortized $O(\log{\Delta})$ time.
\end{proof}

\subsection*{Deletion} 
\vspace{-0.1cm}
Let $p$ and $q$ be the two points deleted from $P$.
We describe the procedure for deleting $p$. The same procedure is applied for deleting $q$.
Let $Cell_{i}(p)$ be the cell containing $p$ in $\Pi_i$ and let $S_i(p)$ be the set containing $Cell_i(p)$ in $DSS_i$.
For each grid $\Pi_i$, we change the parity of $S_i(p)$ in $DSS_i$.
Then, we find the smallest $i$ such that, in $\Pi_i$, $Cell_i(p)$ contains no other points of $P$ than $p$.
%, and removing $Cell_{i}(p)$ disconnects the component. 
If no such $\Pi_i$ exists, then we do not make any change. 
If all the adjacent cells of $Cell_i(p)$ are empty, then we just remove $S_i(p)$ from $DSS_i$. 
Otherwise, we check whether removing $Cell_{i}(p)$ disconnects the component containing it. 
That is, we check whether there are two non-empty adjacent cells of $Cell_{i}(p)$ that were in the same set $S_i(p)$ together with $Cell_{i}(p)$ in $DSS_i$ and after removing $Cell_i(p)$ they should be in different sets. If there are two such cells, then we remove the set $S_i(p)$ from $DSS_i$ and reconstruct new sets for the cells in $S_i(p)\setminus \{Cell_i(p)\}$. 

\begin{lemma}
There is at most one grid $\Pi_i$, such that removing $Cell_{i}(p)$ disconnects the component containing it in $DSS_i$.
%$Cell_i(p) \cap P = \{p\}$, i.e., $Cell_i(p)$ contains no other points of $P$ than $p$. % .
\end{lemma}
\begin{proof}
Assume by contradiction that there are two grids $\Pi_i$ and $\Pi_j$, such that $i < j$ and removing $Cell_{i}(p)$ and $Cell_{j}(p)$ disconnect the component containing it in $DSS_i$ and in $DSS_j$, respectively.
Let $\sigma_1$ and $\sigma_2$ be two non-empty adjacent cells of $Cell_{i}(p)$ in $\Pi_i$ that were in the same set $S_i(p)$ together with $Cell_{i}(p)$ in $DSS_i$.
Notice that $\sigma_1$ and $\sigma_2$ are contained in the $3 \times 3$ cells-block around $Cell_i(p)$ in $\Pi_i$; see Figure~\ref{fig:grid}.
Moreover, one of the corners of $Cell_i(p)$ is a grid-vertex in $\Pi_{i+1}$, as depicted in Figure~\ref{fig:grid}.
Therefore, $\sigma_1$ and $\sigma_2$ are either in the same cell or in adjacent cells in $\Pi_{i+1}$, and in $\Pi_j$, for each $j\ge i+1$.
This contradicts that $Cell_{j}(p)$ disconnects the component containing it in $DSS_j$.
\begin{figure}[htb]
    \centering
    \includegraphics[width=0.9\textwidth]{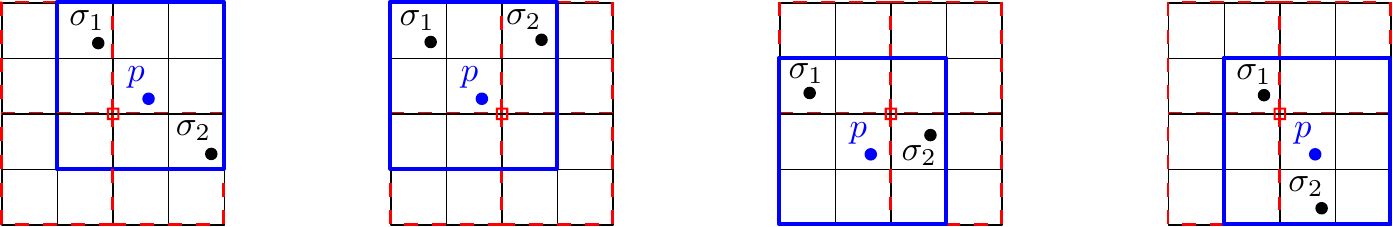}
    \caption{$\Pi_i$ (in black) and $\Pi_{i+1}$ (in red). The $3 \times 3$ cells-block (in blue) around $Cell_i(p)$ in $\Pi_i$. One of the corners of $Cell_i(p)$ is a grid-vertex in $\Pi_{i+1}$.}
    \label{fig:grid}
\end{figure}
\end{proof}

%\vspace{-0.3cm}
\begin{lemma}
Deleting $p$ from $P$ takes amortized $O(\log{\Delta} + |P|)$ time.
\end{lemma}
\begin{proof}
Changing the parity of $S_i(p)$ in $DSS_i$ can be done in constant time, for each $1 \le i \le c$. 
Finding the smallest $i$ such that $Cell_i(p)$ contains no other points of $P$ than $p$ takes $O(\log{\Delta})$ time.
If all the adjacent cells of $Cell_i(p)$ are empty, then we just remove $S_i(p)$ from $DSS_i$ in constant time.
Otherwise, reconstruct new sets for the cells in amortized $S_i(p)\setminus \{Cell_i(p)\}$ in $O(|S_i(p)|) = O(|P|)$ time.
Since $c=\ceil{\log \Delta}$, Deleting $p$ from $P$ takes amortized $O(\log{\Delta} + |P|)$ time.
\end{proof}

The following theorem summarizes the result of this section.
\begin{theorem} \label{thm:Theorem2}
Let $P$ be a set of points in the plane and let $\Delta$ be the geometric spread of $P$.
There exists a dynamic algorithm that maintains a $(6\sqrt{2})$-approximate bottleneck matching of $P_k$, at each time step $k$, and supports insertion in $O(\log{\Delta})$ amortized time and deletion in $O(\log{\Delta} + |P_k|)$ amortized time. 
\end{theorem}

%\newpage
%%%%%%%%%%%%%%%%%%%%%%%%%%%%%%%%%%%%%%%%%%%%%%%%%%%%%%%%%%%%%%%%%%%%%%%%%%%%%%
%%%%%%%%%%%%%%%%%%%%%%%%%%%%%%%%%%%%%%%%%%%%%%%%%%%%%%%%%%%%%%%%%%%%%%%%%%%%%%
%\bibliographystyle{plain}
\bibliography{arxiv}

%%%%%%%%%%%%%%%%%%%%%%%%%%%%%%%%%%%%%%%%%%%%%%%%%%%%%%%%%%%%%%%%%%%%%%%%%%%%%%%%%%%
%%%%%%%%%%%%%%%%%%%%%%%%%%%%%%%%%%%%%%%%%%%%%%%%%%%%%%%%%%%%%%%%%%%%%%%%%%%%%%%%%%%%

\end{document}